# Critical exponents and amplitudes of analytical equation of state

Ikhtier H. Umirzakov [*]

*Laboratory of modeling, Institute of Thermophysics, Lavrenteva prospect, 1, Novosibirsk, 630090, Russia*

The paper analyzes a general case of an equation of state, which is an analytical function at the critical point of the liquid-vapor first order phase transition of pure substance. It is shown that the equality to zero of the first- and second-order partial derivatives of pressure with respect to volume (density) at the critical point is the consequence of the thermodynamic conditions of phase equilibrium. We obtained the relations of critical exponents and amplitudes with parameters of the analytical equation of state. It is shown that the substance with the analytical equation of state can have critical exponents of lattice gas which is equivalent to the two dimensional Ising model. It is shown that the analytical equation of state can take into account the density fluctuations.

The equation of state (the dependence of the pressure $p(T,\rho)$ on the temperature $T$ and the density $\rho$) plays a significant role in the equilibrium thermodynamics of fluids [1-8]. The properties of fluids at a critical point of the liquid-vapor first order phase transition are of a great interest [1-9]. The large number of equations of state (EOS) used in the chemical engineering is an analytical function of temperature and density [1]. Therefore, it is interesting and actual to study a general form of the analytical equation of state at a critical point of the liquid-vapor first order phase transition of pure (one-component) substance.

For further convenience in this work, we use, instead of density $\rho$, the volume per particle (atom or molecule) $v$, which is defined from relation $v = m/\rho$, where $m$ is the mass of one particle.

An analytical equation of state at the critical point being the Taylor series [10] is

$$p(T,v) = p_c + \sum_{k=1}^{\infty} a_k \cdot (T_c - T)^k + \sum_{l=1}^{\infty}\sum_{m=1}^{\infty} b_{l,m} \cdot (T_c - T)^l \cdot (v - v_c)^m + \sum_{n=1}^{\infty} c_n \cdot (v - v_c)^n, \quad (1)$$

where $T_c$ is the critical temperature, $v_c$ is the critical volume per particle, $p_c = p(T_c, v_c)$ is the critical pressure, and

$$a_k = \frac{1}{k!} \cdot \left.\frac{\partial^k p(T,v)}{\partial T^k}\right|_{T_c,v_c}, \quad b_{l,m} = \frac{1}{l!m!} \cdot \left.\frac{\partial^{l+m} p(T,v)}{\partial T^l \partial v^m}\right|_{T_c,v_c}, \quad c_n = \frac{1}{n!} \cdot \left.\frac{\partial^n p(T,v)}{\partial v^n}\right|_{T_c,v_c}.$$

Assume that near critical point the terms $a_K \cdot (T_c - T)^K$, $b_{L,M} \cdot (T_c - T)^L \cdot (v - v_c)^M$ and $c_N \cdot (v - v_c)^N$ in EOS (1) are the main terms of three corresponding sums, so

$$a_k = 0, k = 0,...,K-1, \ a_K \neq 0, \ K \geq 1,$$
$$b_{l,m} = 0, l = 1,...,L, \ m = 1,...,M, \ (l,m) \neq (L,M), \ b_{L,M} \neq 0, \ L \geq 1, \ M \geq 1, \quad (2)$$
$$c_n = 0, n = 0,...,N-1, \ c_N \neq 0, \ N \geq 1,$$

$$p(T,v) = p_c + a_K \cdot (T_c - T)^K + b_{L,M} \cdot (T_c - T)^L \cdot (v - v_c)^M + c_N \cdot (v - v_c)^N. \quad (3)$$

Consider the liquid-vapor phase equilibrium, so $T < T_c$. Using EOS (3) and the condition of equality [3] of temperatures and pressures of liquid and vapor, coexisting in phase equilibrium with each other, we obtain

$$b_{L,M} \cdot (T_c - T)^L \cdot (v_L - v_c)^M + c_N \cdot (v_L - v_c)^N = b_{L,M} \cdot (T_c - T)^L \cdot (v_G - v_c)^M + c_N \cdot (v_G - v_c)^N, \quad (4)$$

where $v_L = v_L(T)$ and $v_G = v_G(T)$ are the volumes per particle in liquid and vapor, coexisting in phase equilibrium, respectively.

The condition of equality of chemical potentials of liquid and vapor coexisting in phase equilibrium with each other gives [3]

$$\int_{v_L}^{v_G} p(T,v) dv = p_e \cdot (v_G - v_L), \quad (5)$$

where

$$p_e = p_e(T) = p(T, v_L) = p(T, v_G) \quad (6)$$

is the pressure of liquid and vapor, coexisting in the phase equilibrium. Using (5)-(6), we have

$$\int_{v_L}^{v_G} p(T,v) dv = p(T, v_G) \cdot (v_G - v_c) - p(T, v_L) \cdot (v_L - v_c). \quad (7)$$

Using (3)-(5) and (7), we obtain

$$b_{L,M} M (T_c - T)^L \cdot \frac{(v_G - v_c)^{M+1} - (v_L - v_c)^{M+1}}{M+1} + c_N N \cdot \frac{(v_G - v_c)^{N+1} - (v_L - v_c)^{N+1}}{N+1} = 0. \quad (8)$$

Particularly, one can easily see that the conditions of the liquid-vapor phase equilibrium (4) and (8) take place if

$$v_G - v_c = -(v_L - v_c), \quad (9)$$

and $M$ and $N$ are odd numbers.
From (4) one can obtain

$$c_N \cdot (v_G - v_c)^{N-M} = -b_{L,M} \cdot (T_c - T)^L. \quad (10)$$

Return to the general case. From (4) and (8) we have

$$\frac{M}{M+1} \cdot \frac{1 - x^{M+1}}{1 - x^M} = \frac{N}{N+1} \cdot \frac{1 - x^{N+1}}{1 - x^N}, \quad (11)$$

where $x = (v_G - v_c)/(v_L - v_c)$. We have $v_G - v_c > 0$ and $v_L - v_c < 0$ from physical conditions, so $x < 0$. One can see graphically or analytically that $x = -1$, corresponding to (9), is the single negative root of the equation (11) if $M$ and $N$ are odd numbers.

Later we assume that $M$ and $N$ are odd numbers.

Using the physical condition $v_G \to v_c +$ when $T \to T_c -$ and the condition $L \geq 1$ (see (2)) from (10) we have $N - M > 0$ and

$$v_G - v_c = (-b_{L,M}/c_N)^{\frac{1}{N-M}} \cdot (T_c - T)^{\frac{L}{N-M}} = A_\beta \cdot (T_c - T)^\beta, \quad (12)$$

$$\beta = \frac{L}{N-M} > 0, \qquad (13)$$

$$A_\beta = (-b_{L,M}/c_N)^{\frac{1}{N-M}},$$

$$b_{L,M}/c_N < 0, \qquad (14)$$

where $\beta$ and $A_\beta$ are the critical exponent (or index) and critical amplitude of the liquid-vapor coexisting line, respectively [3].

From the conditions $M \geq 1$ (see (2)), $M$ and $N$ are odd numbers, and $N - M > 0$ we have

$$N \geq 3. \qquad (15)$$

From (3) we have

$$p(T_c, v) - p_c = c_N \cdot (v - v_c)^N \qquad (16)$$

for the critical isotherm.

Using (15) and (16), we have

$$\partial p(T,v)/\partial v \big|_{T_c, v_c} = 0, \qquad (17)$$

$$\partial^2 p(T,v)/\partial v^2 \big|_{T_c, v_c} = 0. \qquad (18)$$

Note that we have obtained the conditions (17) and (18) from the conditions of the phase equilibrium (4) and (8) without using the condition of thermodynamic stability of substance at the critical point.

Usually, the conditions (17) and (18) are considered as conditions of thermodynamic stability of substance at the critical point [2].

Note the condition (17) was obtained in [3] from the condition (4), and the condition (18) was obtained from the condition of thermodynamic stability of substance at the critical point [3].

Using $p(T_c, v) - p_c = A_\delta \cdot (v - v_c)^\delta$ [3], for the critical exponent $\delta$ and critical amplitude $A_\delta$ of the critical isotherm we have from (16) the following relations

$$\delta = N, \qquad (19)$$

$$A_\delta = c_N.$$

Usually the pressure $p$ decreases along with increasing volume $v$ at the critical isotherm [1,3,7,8], so from (16) we have for this case

$$c_N < 0. \qquad (20)$$

From (14) and (20) we have

$$b_{L,M} > 0. \qquad (21)$$

Usually, $\partial p(T,v)/\partial T \big|_{T_c, v_c} > 0$ [1,3,7,8]. We have from (3) for this case

$K = 1$,
$a_K < 0$.

We have from (3)

$$\partial p(T,v)/\partial v = b_{L,M} M \cdot (T_c - T)^L \cdot (v - v_c)^{M-1} + c_N N \cdot (v - v_c)^{N-1}. \tag{22}$$

The inequality $\partial p(T,v)/\partial v < 0$ takes place for any $v$ at $T > T_c$, particularly, for $|T_c - T|^L \gg |c_N N / b_{L,M} M| \cdot (v - v_c)^{N-M}$ [1,3,7,8]. Therefore, we have from (21) and (22) that $L$ is odd number.

We have for an isothermal compressibility $\kappa(T)$ from (12) and (22)

$$\kappa(T) \equiv -\frac{1}{v_c \partial p(T,v)/\partial v\big|_{v=v_G \text{ or } v=v_L}} = \frac{(T_c - T)^{-\frac{L}{N-M}\cdot(N-1)} \cdot (-c_N)^{\frac{M-1}{N-M}}}{v_c \cdot (b_{L,M})^{\frac{N-1}{N-M}} \cdot (N-M)} = A_\gamma \cdot (T_c - T)^{-\gamma}. \tag{23}$$

From (23) we have for the critical exponent $\gamma$ and critical amplitude $A_\gamma$ of the isothermal compressibility the following relations

$$\gamma = \frac{L}{N-M} \cdot (N-1), \tag{24}$$

$$A_\gamma = \frac{(-c_N)^{\frac{M-1}{N-M}}}{v_c \cdot (b_{L,M})^{\frac{N-1}{N-M}} \cdot (N-M)}.$$

One can see that the equality of Widom [11]

$$\gamma = \beta \cdot (\delta - 1)$$

can be obtained from (13), (19) and (24).

One can see from (24) that $\gamma = L$ for arbitrary odd numbers $L \geq 1$ and $N \geq 3$, if $M = 1$. Particularly, there is the case $L = 1$, $M = 1$, $N = 5$, $\delta = 5$, $\beta = 1/4$, and $\gamma = 1$.

The simplest case corresponds to $L = 1$, $M = 1$, $N = 3$, $\delta = 3$, $\beta = 1/2$ and $\gamma = 1$. It corresponds to the Van-der-Waals theory of critical point and general case of mean field theory [3,5].

The case $L = 1$, $M = 7$, $N = 15$, $\delta = 15$, $\beta = 1/8$ and $\gamma = 7/4$ corresponds to the two-dimensional Ising model with the nearest neighbor interactions [3,5] which is equivalent to lattice gas [5,6].

According to Van der Waals theory of the critical point [3], the Taylor series (1) can be used if no fluctuations are taken into account. If it is true, then the lattice gas does not take into account the fluctuations. However, according to [3,5,6] the Ising model takes into account the fluctuations. So, there is an intrinsic contradiction in the Van der Waals theory of the critical point [3]. We have shown that the substance with the equation of state which is analytical at the critical point can have critical exponents of lattice gas.

Note that the case, when $\beta \equiv 1/3$ exactly, cannot be obtained from the analytical equation of state at the critical point because the equality $N - M = 3L \, (= L\beta^{-1})$ can never be true for odd numbers $L, M$, and $N$ because $3L$ is odd number and $N - M$ is even number. But if $\beta \approx 1/3$, then one can find the values of odd numbers $L, M$, and $N$ which give the approximate quantity of $\beta$. For example, for $\beta = 0.345 = 345/1000 = 69/200$ we have $N - M = 200$ and $L = 69$.

The results presented above were obtained using a general assumption of analyticity of the equation of state and without using any theory (model) such as the approximate fluctuation theory of the critical point [3]. However, the results obtained in present paper do not contradict this theory and other physical models.

The volume of the system $V$ and the total number of particles $N$ in the system are constants in the microcanonical and canonical ensembles [4]: $V = const$ and $N = const$, so the density $\rho$, which is equal to $\rho = mN/V$, has the constant value for the microcanonical and canonical ensembles. It is evidently that the same is true in the thermodynamic limit $V \to \infty$, $N \to \infty$ and $N/V = const$. So, there are no fluctuations of the density of the system as a whole in these ensembles. If the properties of pure substance at and near critical point are related with density fluctuations of the system as a whole, which can take place in the grand canonical ensemble [4], then the microcanonical and canonical ensembles are not equivalent to the grand canonical one. So, the proof of the equivalence of three ensembles [4] is not valid in this case.

This paper considered a general case of the equation of state, which is an analytical function at the critical point of the liquid-vapor first order phase transition of pure substance. The relations of critical exponents and amplitudes with the parameters of the analytical equation of state are obtained.

We have obtained the following results. It is shown that:
- the equality to zero of the first- and second-order partial derivatives of pressure with respect to volume (density) at the critical point is the consequence of the thermodynamic conditions of phase equilibrium;
- the substance with the analytical equation of state can have critical exponents of lattice gas which is equivalent to the two dimensional Ising model; and
- the analytical equation of state can take into account density fluctuations.

[*] e-mail: umirzakov@itp.nsc.ru